\documentstyle[prd,aps,preprint,epsfig]{revtex}
\begin{document}
\newcommand{\TeV}{\,{\rm TeV}}
\newcommand{\GeV}{\,{\rm GeV}}
\newcommand{\MeV}{\,{\rm MeV}}
\newcommand{\keV}{\,{\rm keV}}
\newcommand{\eV}{\,{\rm eV}}
\def\ap{\approx}
\def\bea{\begin{eqnarray}}
\def\eea{\end{eqnarray}}
\def\bec{\begin{center}}
\def\ec{\end{center}}
\def\pC{\tilde{\chi}^+}
\def\nC{\tilde{\chi}^-}
\def\pnC{\tilde{\chi}^{\pm}}
\def\Ne{\tilde{\chi}^0}
\def\snu{\tilde{\nu}}
\def\tN{\tilde N}
\def\ler{\lesssim}
\def\gtr{\gtrsim}
\def\beq{\begin{equation}}
\def\eeq{\end{equation}}
\def\haf{\frac{1}{2}}
\def\plb#1#2#3#4{#1, Phys. Lett. {\bf #2B} (#4) #3}
\def\plbb#1#2#3#4{#1 Phys. Lett. {\bf #2B} (#4) #3}
\def\npb#1#2#3#4{#1, Nucl. Phys. {\bf B#2} (#4) #3}
\def\prd#1#2#3#4{#1, Phys. Rev. {\bf D#2} (#4) #3}
\def\prl#1#2#3#4{#1, Phys. Rev. Lett. {\bf #2} (#4) #3}
\def\mpl#1#2#3#4{#1, Mod. Phys. Lett. {\bf A#2} (#4) #3}
\def\rep#1#2#3#4{#1, Phys. Rep. {\bf #2} (#4) #3}
\def\lpp{\lambda''}
\def\ccg{\cal G}
\def\slash#1{#1\!\!\!\!\!/}
\def\rpv{\slash{R_p}}
\def\ler{\lesssim}
\def\gtr{\gtrsim}
\def\beq{\begin{equation}}
\def\eeq{\end{equation}}
\def\haf{\frac{1}{2}}
\def\plb#1#2#3#4{#1, Phys. Lett. {\bf #2B} (#4) #3}
\def\plbb#1#2#3#4{#1 Phys. Lett. {\bf #2B} (#4) #3}
\def\npb#1#2#3#4{#1, Nucl. Phys. {\bf B#2} (#4) #3}
\def\prd#1#2#3#4{#1, Phys. Rev. {\bf D#2} (#4) #3}
\def\prl#1#2#3#4{#1, Phys. Rev. Lett. {\bf #2} (#4) #3}
\def\mpl#1#2#3#4{#1, Mod. Phys. Lett. {\bf A#2} (#4) #3}
\def\rep#1#2#3#4{#1, Phys. Rep. {\bf #2} (#4) #3}
\def\lpp{\lambda''}
\def\ccg{\cal G}
\def\slash#1{#1\!\!\!\!\!/}
\def\rpv{\slash{R_p}}
\def\pslash{p\hspace{-2.0mm}/}
\def\qslash{q\hspace{-2.0mm}/}
\newcommand{\imag}{\Im {\rm m}}
\newcommand{\real}{\Re {\rm e}}

\setcounter{page}{1}
\draft
\preprint{KAIST-TH 02/16, KIAS-P02038}
\title{Radius-dependent gauge unification in AdS5}
\author{$^{(a)}$Kiwoon Choi,
$^{(b)}$Hyung Do Kim
and $^{(a)}$Ian-Woo Kim}
%
\address{$^{(a)}$Department of Physics,
Korea Advanced Institute of Science and Technology\\ Daejeon
305-701, Korea\\ $^{(b)}$School of Physics, Korea Institute for
Advanced Study, Cheongryangri-dong, Dongdaemun-ku, Seoul 135-012,
Korea}

\date{\today}
%
%
\maketitle
%
\begin{abstract}
We examine the relation of the 4-dimensional low energy coupling
of bulk gauge boson in a slice of AdS5 to the 5-dimensional fundamental couplings
as a function of the orbifold radius $R$.
This allows us to address the gauge coupling unification in AdS5 by means of
the radius running as well as the conventional momentum running.
We then compute the radius dependence of 1-loop low energy couplings
in generic AdS5 theory with 4-dimensional supersymmetry, and
discuss the low energy predictions when the 5-dimensional
couplings are assumed to be unified.

\end{abstract}
%

It has been noted that the large scale hierarchy between the weak
and Planck scales can be naturally obtained in 5-dimensional (5D)
theory on a slice of AdS5 \cite{Randall:1999ee} with an appropriate
radion stabilization mechanism \cite{goldberger}. 
In the original
model of Randall and Sundrum, all the standard model fields are
assumed to be confined in the TeV brane. An apparent drawback of
this scenario is that one has to abandon the perturbative
unification of gauge couplings at the fundamental scale of the
model. An alternative scenario which may achieve gauge unification
while  solving the hierarchy problem is that the gauge fields
propagate in 5D bulk spacetime \cite{Chang:1999nh}.
In such case, the size of
gauge coupling renormalization is generically of order
$[\ln(M_{Pl}/M_W)^2]/8\pi^2$
\cite{Pomarol:2000hp,Randall:2001gb,Choi:2002wx,Goldberger:2002cz,Agashe:2002bx}, 
so the gauge unification can be achieved at a scale
near the 4D Planck scale $M_{Pl}$. In this paper, we first point out
that it is convenient to consider the orbifold radius {\it
$R$-dependence} of 4D couplings in addition to the {\it
momentum-dependence} in order to address the unification of bulk
gauge couplings in AdS5. We then compute the $R$-dependence of
1-loop 4D couplings in generic AdS5 theory with $N=1$
supersymmetry (SUSY) which is orbifolded by $Z_2\times Z^\prime_2$
\cite{Altendorfer:2000rr}, and examine the
low energy consequences of unified 5D couplings.

Let us consider 5D gauge theory on a slice of AdS5 with orbifold radius
$R$. The action includes
\beq
\label{5daction}
\int d^4xdy \sqrt{-G}\,\left(\,-\frac{1}{4g_{5a}^2}F^{aMN}F^a_{MN}
 -\sum_i\frac{\delta(y-n_i\pi)}{\sqrt{G_{55}}}
\frac{1}{4g_{ia}^2}F^{a\mu\nu}F^a_{\mu\nu}\,\right)\,,
\eeq
where  $y=n_i\pi $ $(n_i=0,1)$ denote the 5-th coordinates
of orbifold fixed points and
the 5D metric $G_{MN}$ is given by
\beq
\label{5dmetric}
ds^2=G_{MN}dx^Mdx^N=e^{-2kR|y|}g_{\mu\nu}dx^\mu dx^\nu+R^2dy^2\,,
\eeq
where $k$ is the AdS curvature.
For generic bulk fields in AdS5, Kaluza-Klein (KK) scale
is set by $M_{KK}\simeq {\pi k}/(e^{\pi k R}-1)$
\cite{Chang:1999nh}. 
The quantity of our interest is the 1-loop low energy coupling of 
zero mode gauge boson
with external 4D momentum $p\lesssim M_{KK}$ for generic value of $kR$:
\beq
\label{4dcoupling}
\frac{1}{g_a^2(p,R,k)}\,=\,
\left[\,\frac{1}{g_{5a}^2(\Lambda)}
+\frac{\gamma_a}{8\pi^3}\Lambda\,\right]\pi R
+\left[\sum_i\frac{1}{g_{ia}^2(\Lambda)}+\frac{b^{\prime}_a}{8\pi^2}\ln \Lambda
\right]+\frac{1}{8\pi^2}\tilde{\Delta}_a(p, R, k)\,,
\eeq
where $\Lambda$ is the cutoff scale measured by $G_{MN}$ and
$p^2=-g^{\mu\nu}\partial_\mu\partial_\nu$ is measured by $g_{\mu\nu}$.
Here the linear divergence is from the bulk counter term \cite{dienes},
while the log divergence is from the fixed point counter terms
\cite{Contino:2001si,Hall:2001xb}, 
and  the conventional momentum-running and finite KK threshold corrections
are encoded in $\tilde{\Delta}_a$.
In this paper, we focus  on
\beq
\label{1loop}
\Delta_a=
b^{\prime}_a\ln\Lambda+\tilde{\Delta}_a
\eeq 
which is {\it unambiguously calculable} within 5D effective field theory, 
{\it not} on the uncalculable bare parameters
$1/g_{ia}^2$ and 
$1/{\hat{g}_{5a}^2}\equiv 1/{g_{5a}^2}+\gamma_a \Lambda/8\pi^3$.
In fact, $1/g_{ia}^2$ can be simply ignored 
under the strong coupling assumption 
$g_{ia}^2(\Lambda)={\cal O}(8\pi^2)$
\cite{Chacko:1999hg}.

Let us first summarize some generic features of the 1-loop correction (\ref{1loop}).
Since the UV divergence structure is independent of $k$,
$b^{\prime}_a$ can be computed in the flat limit $k=0$, yielding
\cite{Contino:2001si}
\beq
\label{logdivergence}
b^{\prime}_a=
\frac{1}{12}\left[\,\sum T_a(\phi^{++})-\sum T_a(\phi^{--})
-23\sum T_a(A_M^{++})+23\sum T_a(A_M^{--})\,\right],
\eeq
where $T_a(\Phi)={\rm Tr}(T_a^2)$. 
Here $\phi^{zz^\prime}$ and $A_M^{zz^\prime}$ ($z,z^\prime=\pm$)
are the 5D real scalar and vector fields
with the boundary conditions
$\Phi^{zz^\prime}(-y)=z\Phi^{zz^\prime}(y)$,
$\Phi^{zz^\prime}(-y+\pi)=z^\prime\Phi^{zz^\prime}(y+\pi)$.
When $p\lesssim M_{KK}$, the $p$-dependence of $1/g_a^2$ is given by
${\partial\Delta_a}/{\partial\ln p}=-b_a+{\cal O}(p^2/M_{KK}^2)$
where $b_a$ is the conventional 1-loop beta function coefficient 
in 4D effective theory.
Also for $p\ll 1/R\ll \Lambda$ with $k=0$, one finds
\beq
\label{rdependence}
(\Delta_a)_{k=0}=
b^{\prime}_a\ln (\Lambda R) -b_a\ln(pR)+{\cal O}(1).
\eeq

In large radius limit $\pi kR\gg 1$, the $p$-dependence of $1/g_a^2$
can be determined within 5D field theory {\it only} for $p\lesssim e^{-k\pi R}\Lambda$.
This can be easily seen for instance by considering the effects of higher-derivative terms
in 5D lagrangian density, e.g. $F^{aMN}G^{PQ}\partial_P\partial_QF^a_{MN}/\Lambda$.
Such term  gives a contribution of order $p^2/e^{-2k\pi R}k\Lambda$ to
$1/g_a^2$, which means that 5D field theory description {\it breaks down} for
$p\gtrsim e^{-k\pi R}\sqrt{k\Lambda}$ \cite{Goldberger:2002cz}.
So one can not probe a possible gauge unification
at $\Lambda$ by means of the momentum-running alone.
Physically, this is to be expected since the gauge field zero mode is constant along
$y$, so its amplitude receives an important contribution
from $y\approx\pi$ which has the cutoff $\sim e^{-\pi kR}\Lambda$.
On the other hand, in small radius limit $\pi kR\lesssim 1$,
the leading $p$-dependence  is calculable within
the 5D field theory of (\ref{5daction}) as long as 
$p\lesssim {\cal O}(\Lambda)$.
In particular, when $k\approx 1/\pi R\approx \Lambda$, we have
$\left(\Delta_a\right)_{k\approx 1/\pi R\approx \Lambda}\approx
b_a \ln({\Lambda}/{p})$. 

With the above observation, the gauge coupling renormalization in AdS5 
can be described by Fig.1.
First of all, the range of $[\,\ln(p/M_{Pl}), \pi kR\,]$ which
allows a 5D field theory description  is bounded to be below
the line A representing $\ln(p/M_{Pl})\approx -\pi kR$.
The momentum-running of $g_a^2$ when $\pi kR\gg 1$
is allowed only for $p\lesssim e^{-\pi kR}\Lambda$ (the line 1), while
for $\pi kR\lesssim 1$,
the momentum-running is allowed up to $p={\cal O}(\Lambda)$ (the line 3).
At $p\approx\Lambda\approx 1/\pi R\approx k$, we have $1/g_a^2\approx
\pi R/\hat{g}^2_{5a}$.
From this point, one can move along the dotted lines 
to arrive at the phenomenologically relevant point
with $p\approx M_W$ and $\pi kR\gg 1$.
This procedure involves always a {\it radius-running}
along the line 2, so 
it is crucial to compute the $R$-dependence of $g_a^2$ over 
the range from $\pi kR\lesssim 1$
to $\pi kR\gg 1$ in order to 
determine $g_a^2$ at $p\approx M_W$ and $\pi kR\gg 1$.
This suggests  also that the gauge unification in AdS5 can be addressed 
by means of the double running along $\ln p$ and $R$.
Suppose that the 5D bare couplings $\hat{g}_{5a}^2$ at $\Lambda$ have a 
unified value.
Then the resulting prediction on $g_a^2$ 
at $p\approx M_W$ and $\pi kR\gg 1$
can be unambiguously computed by means of the $\ln p$ and 
$R$ runnings.

It is possible to directly compute the
$R$-dependence of $g_a^2$ in generic AdS5 theory \cite{choi-kim}. 
However in
case with unbroken $N=1$ SUSY, there is much simpler way to
compute the $R$-dependence. In SUSY case, the radion $R$ forms a
$N=1$ superfield together with the 5-th
component of the graviphoton $B_M$. In 4D effective supergravity
(SUGRA), the field-dependence of gauge couplings is determined by
the field-dependence of holomorphic gauge kinetic function $f_a$
and K\"ahler potential $K$ which can be expanded in powers of
generic charged superfield $\Phi$: $K=
K_0(T,T^*)+Z_{\Phi}(T,T^*)\Phi^*e^{-V}\Phi$, where $T$
denotes generic gauge singlet moduli superfield. Then the
moduli-dependence of 1-loop low energy couplings are unambigously
determined to be \cite{Kaplunovsky:1994fg}
\bea
\label{4dsugracoupling} \frac{1}{g^2_a(p)}\,&=&\,
\mbox{Re}(f_a)+\frac{b_a}{16\pi^2}
\ln\left(\frac{M_{Pl}^2}{e^{-K_0/3}p^2}\right) \nonumber \\
&&-\sum_{\Phi}\frac{T_a(\Phi)}{8\pi^2}\ln\left(e^{-K_0/3}Z_{\Phi}
\right) +\frac{T_a({\rm Adj})}{8\pi^2}\ln\left({\rm Re}(f_a)\right), \eea
where $b_a=\sum_\Phi T_a(\Phi)-3T_a({\rm Adj})$ and
$M_{Pl}$ is the Planck scale of
$g_{\mu\nu}$ which defines
$p^2=-g^{\mu\nu}\partial_\mu\partial_\nu$. 
With (\ref{4dsugracoupling}),
one can determine the $R$-dependence of 1-loop couplings in AdS5 
by computing the $R$-dependence of $f_a$ and $K$ in
the corresponding 4D effective SUGRA.
Obviously, (\ref{4dsugracoupling}) indicates that
the 1-loop threshold corrections from massive
KK modes are encoded in $f_a$, while
the 4D field-theoretic loop effects of massless modes
can be determined by the {\it tree-level} forms
of $K$ and $f_a$. 
The $R$-dependent 1-loop $f_a$ appears to be
the most nontrivial part to compute.
However, $f_a$ is a holomorphic function
of the radion superfield $T$ whose scalar component is given by
$T=R+iB_5$,
so its $R$-dependence can be determined by  the $B_5$-dependence
which is much easier to compute.

Let us consider a generic supersymmetric 5D theory
on $S^1/Z_2\times Z_2^{\prime}$ with action
\bea
\label{5daction1}
S = \int d^5x&&\sqrt{-G} \,\left[\frac{M_5^3}{2}\left(\,
{\cal R}-\frac{3}{2}C_{MN}C^{MN}\,\right)
+\frac{1}{\hat{g}_{5a}^2}
\left(\frac{1}{2}D_M\phi^aD^M\phi^a-\frac{1}{4}F^{aMN}F^a_{MN}
\right.\right.\nonumber \\
&&\left.\left.+\frac{i}{2}\bar{\lambda}^{ia}\gamma^MD_M\lambda^a_i\right)
+|D_Mh_I^i|^2+i\bar{\Psi}_I\gamma^MD_M\Psi_I+
ic_Ik\epsilon(y)\bar{\Psi}_I\Psi_I +... \right]
\eea
where ${\cal R}$ is the 5D Ricci scalar,
$C_{MN}=\partial_MB_N-\partial_NB_M$ is the graviphoton field strength,
$\phi^a,A_M^a$ and $\lambda^{ia}$ ($i=1,2$) are 5D scalar, vector and symplectic Majorana
spinors constituting a 5D vector multiplet,
$h_I^i$ and $\Psi_I$  are 5D scalar and Dirac spinor constituting the $I$-th hypermultiplet
with kink mass $c_Ik\epsilon(y)$.
For nonzero $k$, 
$U(1)_R$ is gauged as
\bea
&& D_Mh_I^i =\partial_M h_I^i-i\left(\frac{3}{2}(\sigma_3)^i_j-c_I\delta^i_j\right)
k\epsilon(y)B_M h^j_I +...\nonumber \\
&& D_M\Psi_I=\partial_M \Psi_I +ic_Ik\epsilon(y)B_M\Psi_I+...
\nonumber \\
&& D_M\lambda^{ai}=\partial_M\lambda^{ai}-i\frac{3}{2}(\sigma_3)^i_j k
\epsilon(y)B_M\lambda^{aj}+...\,,
\eea
where the ellipsis stands for other gauge interactions.
The 5D SUGRA multiplet is assumed to have the standard boundary conditions
under $Z_2: y\rightarrow -y$ and $Z_2^{\prime}:y+\pi\rightarrow -y+\pi$,
leaving the 4D $N=1$ SUSY unbroken. 
On the other hand, the vector and hypermultiplets can have arbitrary
boundary conditions:
\bea
\label{bc}
&& A^a_\mu(-y)=z_a A^a_\mu(y),
\quad 
A^a_\mu(-y+\pi)=z^\prime_a A^a_\mu(y+\pi),
\nonumber \\
&& h^i_I(-y)=z_I(\sigma_3)^i_jh^j_I(y),
\quad
h^i_I(-y+\pi)=z^\prime_I(\sigma_3)^i_j h^j_I(y+\pi)\,,
\eea 
where $z_{a,I},z^\prime_{a,I}=\pm 1$ and
the boundary conditions of other 5D fields are fixed by (\ref{bc}) 
in the standard manner. 
To derive the 4D effective SUGRA action, it is convenient to write
the above 5D action in $N=1$ superspace\cite{Arkani-Hamed:2001tb}. 
Among 5D gravity
multiplet, we keep only $T$ and replace
other fields by their vacuum expectation values. Then
following \cite{Arkani-Hamed:2001tb,Marti:2001iw}, we find 
(for $M_5=1$)
\bea \label{5daction2} S_1&=&\int d^5x \,\left[\,\int d^4\theta
\,\left\{\frac{1}{2}(T+T^*)e^{-(T+T^*)k|y|}\left(
1+e^{(\frac{3}{2}-c_I)(T+T^*)k|y|}H_I^*e^{-V}H_I
\right.\right.\right. \nonumber \\
&&\left.\left.+e^{(\frac{3}{2}+c_I)(T+T^*)k|y|}H^c_Ie^VH^{c*}_I)\right)
+\frac{2}{\hat{g}_{5a}^2}\frac{e^{-(T+T^*)k|y|}}{T+T^*}
(\partial_y V^a-\frac{1}{\sqrt{2}}(\chi^a+\chi^{a*}))^2\right\}
\nonumber \\ && \left. +\int d^2\theta\,\left\{\,
\frac{1}{4\hat{g}_{5a}^2}TW^{a\alpha}W^a_{\alpha}
+H^c(\partial_y-\frac{1}{\sqrt{2}}\chi)H +h.c.\,\right\}
\right]\,,\eea
where $W^{a}_{\alpha}$ is the chiral spinor superfield for the
bulk gauge multiplet  $V^a=(A^a_{\mu},\lambda^a)$ with
$\lambda^a=(1-\gamma_5)\lambda^{a1}/2$,
and $H_I=(h_I^1,\psi_I)$,
$H_I^c=(h_I^{2*},\psi^c_I)$, $\chi^a=(\phi^a+iA^a_5, \eta^a)$ are
chiral superfields containing two-component fermions
$\psi_I=(1-\gamma_5)\Psi_I/2$,
$\bar{\psi^c}_I=(1+\gamma_5)\Psi_I/2$,
$\bar{\eta}^a=(1+\gamma_5)\lambda^{a}_1/2$. 
Note that $H_I$ with $z_I=z^\prime_I=1$,
$H^c_I$ with $z_I=z^\prime_I=-1$,
$V^a$ with $z_a=z^\prime_a=1$,  $\chi^a$ with
$z_a=z^\prime_a=-1$ can give massless
4D modes. The model can be easily generalized to include 
$N=1$ superfields $Q_{UV}$ ($Q_{IR}$)
living on the UV (IR) brane at $y=0$ ($\pi$).
In fact, to rewrite (\ref{5daction1}) as (\ref{5daction2}),
one needs to perform  $R$ and $B_5$-dependent field redefinition, 
yielding an additional action through the
chiral anomaly \cite{Arkani-Hamed:2001is}:
\bea
\label{anomaly}
S_2=\int
d^5x d^2\theta
&&\left\{\frac{3}{4}T_a(\lambda^b)
\left(z_b\delta(y)+z^\prime_b\delta(y-\pi)\right)
-\frac{1}{2}c_IT_a(\Psi_I)\left(
z_I\delta(y)+z^\prime_I\delta(y-\pi)\right)\right.
\nonumber \\
&& \left.\,-\frac{3}{2}T_a(\psi_{IR})\delta (y-\pi)\right\}
(16\pi^2)^{-1}{k|y| T}W^{a\alpha}W^a_{\alpha} +h.c.\,,
\eea
where $\psi_{IR}$ is the fermion component of $Q_{IR}$.
Using the {\it holomorphy property}, this anomaly term
can be easily determined  by the following
$B_5$-dependent transformation of fermions:
\bea \label{phase} &&
\lambda^{ai} \rightarrow e^{3ik|y|B_5/2} \lambda^{ai}\,, \,\,
\Psi_I \rightarrow e^{-ic_Ik|y|B_5}\Psi_I\,, \,\,
\psi_{IR}\rightarrow e^{-3ik\pi B_5/2}\psi_{IR}\,.\nonumber 
\eea

$S_1$ has been derived in
\cite{Marti:2001iw} using different superfield
basis for $H_I$ and $H_I^c$. A nice feature of our field basis is
that $B_5$ does {\it not} have any non-derivative interaction in
$S_1$ other than the Chern-Simons coupling. As a result, 
integrating out massive KK modes does not generate 1-loop
$B_5$-coupling to $F\tilde{F}$ other than
those in $S_2$, so {\it no $R$-dependent 1-loop $f_a$ other than those
from $S_2$}. It is then straightforward to derive the
K\"ahler metrics of the massless 4D fields from $H_I,H^c_I$
and of the brane fields $Q_{UV},Q_{IR}$,
and also the gauge kinetic function  of the massless 4D gauge fields
\cite{Marti:2001iw,luty,Bagger:2001}:
\bea \label{kahler} && Y_{H_I,H_I^c}\,=\,
2M_5(e^{(\frac{1}{2}-z_I c_I)\pi k(T+T^*)}-1)/(1-2z_Ic_I)k\,, 
\nonumber \\
&& 
Y_{Q_{IR}}\,=\, e^{-\pi k(T+T^*)}\,,
\quad Y_{Q_{UV}}\,=\, 1\,, \nonumber \\ 
&&f_{a}\,=\,\frac{\pi T}{\hat{g}^2_{5a}} -\frac{3}{8\pi^2}\left(
T_a(\psi_{IR})-\frac{1}{2}\sum_bz^\prime_bT_a(\lambda^b)+\frac{1}{3}
\sum_Iz^\prime_Ic_I
T_a(\Psi_I)\right)k\pi T\,, \eea 
where $Y_{\Phi}=e^{-K_0/3}Z_{\Phi}$ for the
K\"ahler metric $Z_{\Phi}$, and
$M_{Pl}^2=e^{-K_0/3}M_5^2=(1-e^{-k\pi(T+T^*)})M_5^3/k$.
The calculation of $Y_{\chi^b}$ for the massless 4D fields
from $\chi^b$ involves the heavy tadpole with
one 4D derivative, i.e. the tadpole of $A^b_\mu$ with
$z_b=z^\prime_b=-1$ \cite{luty}:
$$
A^b_\mu=\partial_\mu A^b_5\left[\,
y-\pi\frac{e^{(T+T^*)ky}-1}{e^{(T+T^*)k\pi}-1}\right]\,,
$$
yielding
\beq 
\label{kahler1}
Y_{\chi^b} \,=\, k/(e^{\pi k (T+T^*)}-1)M_5.
\eeq
Applying (\ref{kahler}) and (\ref{kahler1}) to (\ref{4dsugracoupling}), 
we find
\bea
\label{4dbulkcoupling}
\Delta_a &=& T_a(Q_{UV})\ln (M_5/p)+T_a(Q_{IR})
\ln(M_5e^{-\pi kR}/p)\nonumber \\
&&-T_a({\rm Adj})\left(\,3\ln(M_5/p) -3\pi k R/2 
-\ln(M_5 R)\,\right)\nonumber \\
&&+\sum_{z_I=z^\prime_I}T_a(H_I)\left(\,\ln(k/p)-z_Ic_I \pi kR 
-\ln \left[(e^{(1-2z_Ic_I)\pi kR}-1)/(\pi(1-2z_Ic_I))\right]\right)
\nonumber \\
&&-\sum_{z_I=-z^\prime_I}z^\prime_IT_a(H_I)c_I \pi kR
+\sum_{z_b=-z^\prime_b}z^\prime_bT_a(V^b) 3\pi kR/2
\nonumber \\
&&+\sum_{z_b=z^\prime_b=-1}T_a(\chi^b) \left(
\ln(M_5^2/pk)+\pi kR/2+\ln(1-e^{-2\pi kR})\right) 
\eea
which is valid for  $p\lesssim M_{KK}\approx \pi k/(e^{\pi kR}-1)$.
The above result obtained by 4D effective SUGRA calculation
can be confirmed by an explicit loop calculation
summing all the loops of KK modes \cite{choi-kim}, which 
assures the validity of our 4D SUGRA calculation.

As a simple example to show the effects of radius-running, let us consider
a supersymetric model with
the MSSM gauge and Higgs superfields living in 5D bulk spacetime
with $z_a=z^\prime_a=z_I=z^\prime_I=1$, 
the MSSM lepton superfields living on 
the UV brane and the MSSM quark superfields on the IR brane.
We choose $k=5\times 10^{17}$ GeV,
$M_5=1.5\times 10^{18}$ GeV, and assume the $SU(5)$-like
boundary conditions of 5D couplings at $M_5$: 
$\hat{g}_{5_{SU(3)}}^2=\hat{g}_{5_{SU(2)}}^2
=5\hat{g}_{5_{U(1)}}^2/3$. We also choose 
$c_I=1/2$ for the Higgs hypermultiplets, so that the 
Higgs zero modes are constant along $y$.
As we have noted, the unification of $\hat{g}_{5a}^2$
implies  that $g_a^2$ are unified at $p\approx 1/R\approx M_5$.
If our universe has $1/R\approx M_5$, 
the 4D effective theory 
for $p\lesssim M_5$ is the MSSM, and then this model can not
be compatible with the observed low energy couplings which indicate that
gauge couplings are unified at $2\times 10^{16}$ GeV, not at $M_5\approx 10^{18}$ GeV.
However if our universe has $\pi kR\approx 10$, the momentum running along 
the line 1 is allowed only for $p\lesssim 7\times 10^{13}$ GeV.  
One then has to include the effects of radius running along the line 2,
making the observed low energy couplings to be consistent with
the 5D unification at $M_5$ as depicted in Fig.2.

To conclude, we have pointed out that the gauge coupling renormalization
in AdS5 can be studied
by means of the double running along $\ln p$ and $R$, as illustrated
in Fig. 1. It is then crucial to compute the $R$-dependence of $g_a^2$
to address the issue of gauge unification in AdS5.  
Using the gauged $U(1)_R$ and chiral anomaly in 5D SUGRA and also
the holomorphic property of 4D effective SUGRA,
we could compute the $R$-dependence of 1-loop 4D couplings in generic AdS5
theory with $N=1$ SUSY which is orbifolded by $Z_2\times Z^\prime_2$.
Our result (\ref{4dbulkcoupling}) can be used to study gauge unification
in supersymmetric AdS5 model.


\begin{figure}
  \centering
  \epsfig{figure=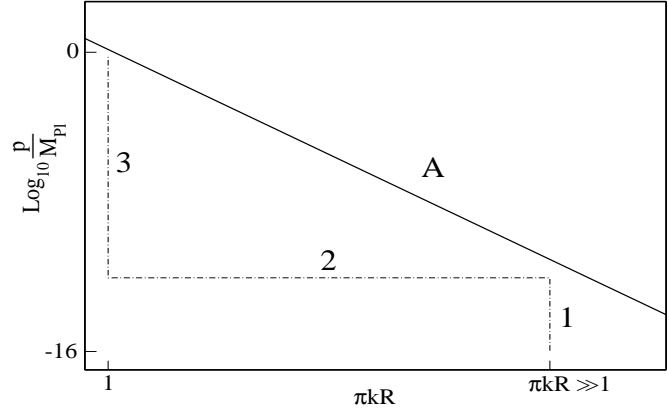, width=10cm, height=7cm , angle=270}
  \caption{The domain of $[\,\ln p,\pi kR\,]$ which can be described
by 5D effective field theory. The line A represents $\ln (p/M_5)\approx -\pi kR$.}
\end{figure}
\begin{figure}
  \centering
  \epsfig{figure=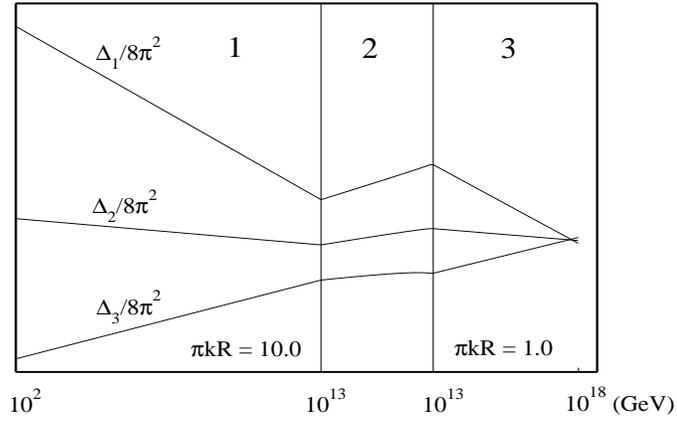, width=10cm, height=7cm , angle=270}
  \caption{The running of 1-loop $\Delta_a$.
The region 1 represents the momentum running 
up to $p\approx 10^{13}$ GeV  along the line 1 of Fig. 1, the 
region 2 is the radius running
from $\pi kR=10$ to $\pi kR=1$ along the line 2, and the
region 3 is the momentum running
to $p\approx M_5$ along the line 3.}
\end{figure}



\begin{thebibliography}{99}
\bibitem{Randall:1999ee}
L.~Randall and R.~Sundrum,
Phys.\ Rev.\ Lett.\  {\bf 83}, 3370 (1999).
\bibitem{goldberger}
W. D. Goldberger and M. B. Wise, Phys. Rev. Lett. {\bf 83}, 4922 (1999).
\bibitem{Chang:1999nh}
S.~Chang, J.~Hisano, H.~Nakano, N.~Okada and M.~Yamaguchi,
Phys.\ Rev.\ D {\bf 62}, 084025 (2000);
T.~Gherghetta and A.~Pomarol,
Nucl.\ Phys.\ B {\bf 586}, 141 (2000);
S.~J.~Huber and Q.~Shafi,
Phys.\ Rev.\ D {\bf 63}, 045010 (2001).
%
\bibitem{Pomarol:2000hp}
A.~Pomarol,
Phys.\ Rev.\ Lett.\  {\bf 85}, 4004 (2000).
\bibitem{Randall:2001gb}
L.~Randall and M.~D.~Schwartz,
JHEP {\bf 0111}, 003 (2001);
Phys.\ Rev.\ Lett.\  {\bf 88}, 081801 (2002).
\bibitem{Choi:2002wx}
K.~Choi, H.~D.~Kim and I.~W.~Kim,
hep-ph/0202257.
\bibitem{Goldberger:2002cz}
W.~D.~Goldberger and I.~Z.~Rothstein,
hep-th/0204160.
\bibitem{Agashe:2002bx}
K.~Agashe, A.~Delgado and R.~Sundrum,
hep-ph/0206099.
\bibitem{Altendorfer:2000rr}
R.~Altendorfer, J.~Bagger and D.~Nemeschansky,
Phys.\ Rev.\ D {\bf 63}, 125025 (2001); 
A.~Falkowski, Z.~Lalak and S.~Pokorski,
Phys.\ Lett.\ B {\bf 491}, 172 (2000);
T. Gherghetta and A. Pomarol, hep-ph/0012378.

\bibitem{dienes} K. R. Dienes, E. Dudas and T. Gherghetta, Phys. Lett. 
B {\bf 436}, 55 (1998);
Nucl. Phys. B {\bf 537}, 47 (1999).
\bibitem{Contino:2001si}
R.~Contino, L.~Pilo, R.~Rattazzi and E.~Trincherini,
Nucl.\ Phys.\ B {\bf 622}, 227 (2002).
\bibitem{Hall:2001xb}
L.~J.~Hall and Y.~Nomura,
Phys.\ Rev.\ D {\bf 65}, 125012 (2002).


%
\bibitem{Chacko:1999hg}
Z.~Chacko, M.~A.~Luty and E.~Ponton,
JHEP {\bf 0007}, 036 (2000);
Y.~Nomura, D.~R.~Smith and N.~Weiner,
Nucl.\ Phys.\ B {\bf 613}, 147 (2001).
%
%
%
%


\bibitem{choi-kim}
K. Choi and I-W. Kim, hep-th/0208071.

\bibitem{Kaplunovsky:1994fg}
V.~Kaplunovsky and J.~Louis,
Nucl.\ Phys.\ B {\bf 422}, 57 (1994).
\bibitem{Arkani-Hamed:2001tb} N.~Arkani-Hamed, T.~Gregoire and J.~Wacker, 
JHEP {\bf 0203}, 055 (2002) 

\bibitem{Marti:2001iw}
D.~Marti and A.~Pomarol,
Phys.\ Rev.\ D {\bf 64}, 105025 (2001).
\bibitem{Arkani-Hamed:2001is}
N.~Arkani-Hamed, A.~G.~Cohen and H.~Georgi,
Phys.\ Lett.\ B {\bf 516}, 395 (2001) 
%
\bibitem{luty} M. A. Luty and R. Sundrum, Phys.\ Rev.\ D {\bf 64}, 065012
(2001). 

\bibitem{Bagger:2001}
J.~Bagger, D.~Nemeschansky and R.~J.~Zhang,
JHEP {\bf 0108}, 057 (2001);
A.~Falkowski, Z.~Lalak and S.~Pokorski,
Nucl.\ Phys.\ B {\bf 613}, 189 (2001).

%


\end{thebibliography}
\end{document}